\documentclass{aa}  
\usepackage{graphicx}
\usepackage{txfonts}
\begin{document}
   \title{Q0045-3337: models including strong lensing by a spiral galaxy}

   %\subtitle{}

   \author{M. Chieregato
          \inst{1,2},
          M. Miranda
          \inst{1}
          \and
          P. Jetzer
          \inst{1}
          }

   \offprints{M. Chieregato}

   \institute{Institut f\"{u}r Theoretische Physik der Universit\"{a}t
     Z\"{u}rich, Winterthurerstrasse 190, CH-8057 Z\"{u}rich, Switzerland\\
\email{matteo@physik.unizh.ch; solar@physik.unizh.ch;\\ jetzer@physik.unizh.ch}
\and
Dipartimento di Fisica e Matematica, Universit\`{a} dell'Insubria,
               Via Valleggio 11, I-22100 Como, Italy\\
}

   \date{Received; accepted}

% \abstract{}{}{}{}{} 
% 5 {} token are mandatory
 
  \abstract
  % context heading (optional)
  % {} leave it empty if necessary  
   {}
  % aims heading (mandatory)
   {Falomo et al. (2005) discovered a disk-like galaxy at $\sim 1.2$
arcsec from the QSO Q0045-3337 by means of ESO VLT adaptive optics. They estimated a galaxy
Einstein radius (for a point mass) of comparable
size, thus pointing up the existence of a new, rare, spiral lens candidate,
despite no evident image splitting.

Here we discuss the possible
lensing effect of the galaxy in some more detail.}
  % methods heading (mandatory)
   {We performed two dimensional
surface photometry on the VLT image of the galaxy, confirming its spiral
nature.  We then
verified if simple mass models, partially constrained by observational data,
require unrealistic parameters to produce a still hidden second quasar
image. We also evaluated the respective viability of an instrumental or a
lensing origin of the observed QSO deformation.} 
  % results heading (mandatory)
   { After galaxy model subtraction, we found a residual
  image, likely not related to gravitational lensing.
Existing data are not sufficient to assess the presence of image
 splitting, nor to constrain the number of images or discriminate between
 galaxy mass models.}
  % conclusions heading (optional), leave it empty if necessary 
   {Further observations are mandatory to progress in the
 study of this remarkable system, that could shed more light on the lensing
 behaviour of spiral galaxies.}


%\titlerunning{Q0045-3337: models including strong lensing by a spiral galaxy}
%\authorrunning{M. Chieregato et al.}   
\maketitle
%
%________________________________________________________________

\section{Introduction}
\label{intro}

The number of known strong lensed quasars boosted in the last years up to the
current value of $\sim 90$ (Kochanek, Schneider \& Wambsganss \cite{diablerets}; CASTLES web
site, http://cfa-www.harvard.edu/castles/). The reason for this increase is
mainly the availability of new observational facilities, in particular the
Hubble Space Telescope, of dedicated surveys (CLASS, Myers et
al. \cite{myers2003}, Browne et al. \cite{browne2003}) and of large public
observational databases, like the FIRST (White et al. \cite{white1997}) and in
particular the Sloan Digital Sky Survey (York et al. \cite{york2000},
Adelman-McCarthy et al. \cite{adel2006}).

  The lion's share of the lens population is made by early-type galaxies.
 In fact,
up to now, only 5 systems are confidently identified with spiral galaxies:
first of all, the Einstein cross, Q2237+0305 (Huchra et
 al. \cite{huchra1985}), then B1600+434 (Jackson et al. \cite{jackson1995}),
PMNJ2004-1349 (Winn, Hall, \& Schechter \cite{winn2003}),
 B0218+357 (Patnaik et al. \cite{patnaik1993}) and PKS 1830-211 (Pramesh Rao
 \& Subrahmanyan \cite{prs1988}).
%A sixth object, the recently discovered CXOCY J220132.8-320144 (Castander et
%al. \cite{castander06}), needs confirmation... Other two strong lenses, B0712+472 (Jackson et al. \cite{jackson1998})
% and APM08279+5255 (Lewis et al. \cite{lewis2002}),    are of uncertain classification, bringing the total to eight objects     
 Each of these systems has its own peculiarities
(inclination of the lens galaxy, richness of the lens environment, etc.), up
to the point of making it unique, and can therefore bring its own precious
contribution to the knowledge of spiral galaxies mass distribution.

The discovery of a late-type galaxy at $\sim 1.2 ''$ from Q0045-3337 (Falomo
 et al. \cite{fal2005}) suggested the existence of another
%seventh spiral lens
 candidate. 

In fact, while there is no evident image splitting, Falomo et al. inferred a galaxy Einstein radius of size comparable to the
 distance between the galaxy and the QSO, under reasonable
hypotheses on galaxy redshift and mass-to-light ratio, and considering the galaxy as a point mass.

Existing, limited data hamper detailed modeling of Q0045-3337. We
 investigated some simple configurations compatible with the known properties
 of the system, in the hope of restricting the range of possibilities and to
 focus future observations.

%__________________________________________________________________

\section{Observational parameters}
\label{observation}
\begin {figure*}[htb]
\centering
  \includegraphics[height=7.62 cm]{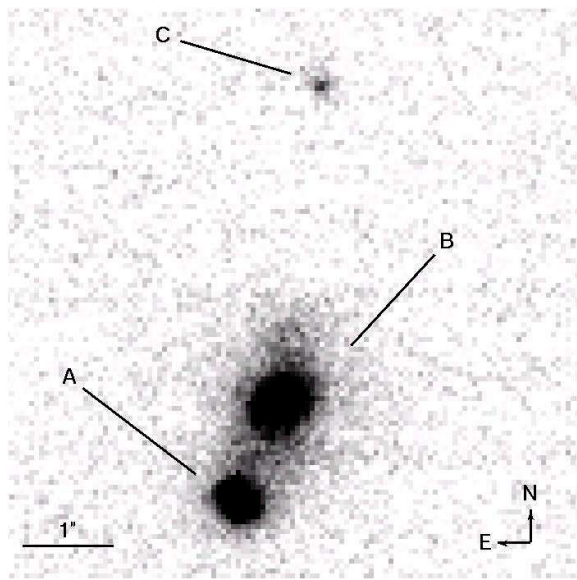}
  
  \caption[Q0045-3337 and its foreground galaxy]{\textbf{Q0045-3337 and its foreground galaxy.} A
  close-up of the NAOS-CONICA image of Q0045-3337 and its foreground
  galaxy.}
  \label{q0045first}
\end {figure*}

%______________________________________________ 
Q0045-3337 (R.A. = 00 47 41.85, DEC = -33 20 55.1) is a radio quiet quasar
with a reported redshift of z=2.14 (Iovino, Clowes \& Shaver
\cite{iovino1996}) and a V magnitude of 18.75 (V\'{e}ron-Cetty \&
V\'{e}ron \cite{veronveron2001}).
The NAOS-CONICA VLT K band image of Q0045-3337 (Falomo et al. \cite{fal2005})
revealed a galaxy at $1.14 ''$ SE from the quasar (respectively A and B in
Figure ~\ref{q0045first}).
 The K magnitude of the QSO is 17.80, while the one of the galaxy
is  16.97 (see the following paragraph for a discussion of the galaxy
surface photometry). The only other object seen in the field, apart from the
  guide star, is source C, with a K magnitude of 20.90,
$3.5 ''$ N of A.

 The quasar itself appears noticeably elliptical
(see Section ~\ref{ellipticity} for a discussion on the ellipticity). 
There is no redshift measurement for the galaxy. Falomo et
al. (\cite{fal2005}) estimated a redshift range of 0.4-1 from an educated
guess on the galaxy absolute K magnitude (-24 to -26) and effective radius (2
to 3 kpc).

\begin {figure*}[htb]
\centering
   \includegraphics[height=7.62 cm]{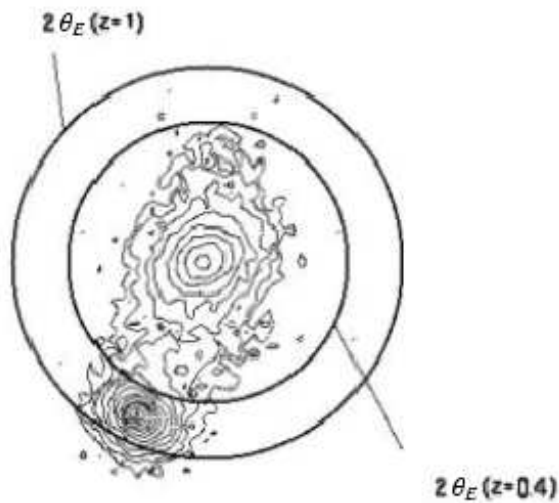}
  
  \caption[Contour map and Einstein radii]{\textbf{Contour map and Einstein radii.} Contour map from the NAOS-CONICA image, with
  superimposed circles of twice the Einstein radius for SIS with dispersion velocities
  calibrated with the Tully-Fisher law (see text for
  details).}
  \label{q0045second}
\end {figure*}

\section[surface photometry and galaxy subtraction]{Surface photometry and galaxy image subtraction}
\label{surface}
We performed two-dimensional surface photometry on the NAOS CONICA image of
the galaxy with the aid of the software
GALFIT (version 2.0.3c; Peng et al. \cite{peng2002}).
Two dimensional photometry for adaptive optics is complicated by the
peculiarities of the Point Spread Function (PSF).
In fact, each extended component of the GALFIT models is usually convolved with the PSF of the image; moreover, the PSF is used to fit point-like
sources.
It is a well known feature of adaptive optics that the PSF can assume complex shapes,
in particular it usually varies with angular distance from the guide star,
and it can be elongated in direction of the guide star.
 Usually, analytical modelling of the PSF from ancillary data
is not possible, though the situation is likely to change in the future (see
Cl{\'e}net et al. \cite{clenet2006}   
for the NaCo case). What should be done is
therefore to infer a PSF template from observations of bright stars at similar
angular distance from the guide star, observations performed just before and
after the one to be analyzed. Unfortunately, this
part of the program was not executed for the NaCo observation of Q0045-3337 (see
Falomo et al. \cite{fal2005}). 

The QSO itself, besides being ``contaminated'' from light
from the galaxy, appears visibly elliptical, and it is not clear if its
ellipticity is an effect of PSF elongation or it is due to a different cause 
(Figure ~\ref{q0045second}; Section ~\ref{ellipticity}). The only other source, object C, is very faint and
irregular, and its point-like or extended nature is not well determined.

We therefore performed a simultaneous fit of the galaxy and the QSO without
convolving a PSF, using GALFIT models also for the QSO light distribution.

A two component model is required to account for the galaxy light
distribution (plus two other components to fit the QSO); in particular, we obtained the best results with two S{\'e}rsic
models of integrated magnitude 17.19 and 18.80, both with S{\'e}rsic index 0.63
(1 is the exponential disk, 0.5 the gaussian) and R$\rm{_e}$ respectively
$13.5 ''$ and $3.4 ''$. The 
inner model is tilted of 50 degrees with respect to the outer.

\begin {figure*}[htb]
\centering
   \includegraphics[height=7.62 cm]{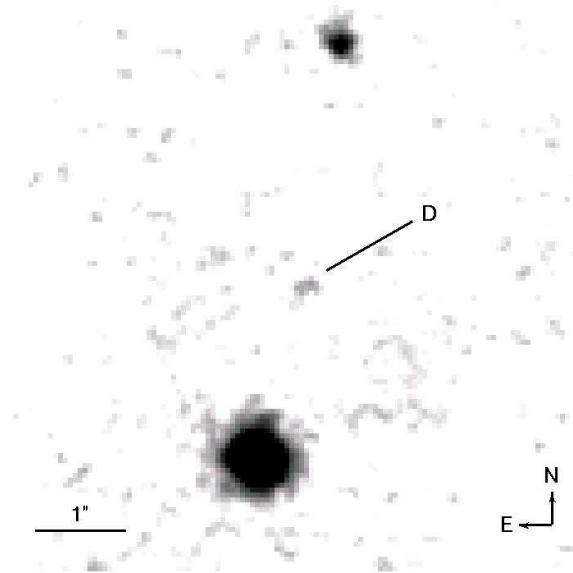}
  
  \caption[Q0045-3337 galaxy subtracted]{\textbf{Q0045-3337 galaxy subtracted.} The NAOS-CONICA image of
  Q0045-3337 after galaxy model subtraction, top-hat smoothing and contrast
  enhancing, in logarithmic scale. The residual feature (D) is well visible midway between object C and QSO.}
  \label{q0045third}
\end {figure*}

After galaxy model subtraction, the main residual feature is an
irregularly shaped object at $0.85 ''$ N of the galaxy centroid ($\sim 2 ''$ from
the QSO; see Figure ~\ref{q0045third}). The  
residual is very elongated, but due to its faintness (K $\sim 22.6$, poorly constrained) it is not possible to
firmly establish its point-like or extended nature. 

%perhaps part of a winded spiral arm poorly fitted by our axisymmetric light distribution
%model, or if it is a point source, either related or unrelated with lensing of
%the QSO. The arclet shape of the residual is somehow...

We checked the robustness of our results using PSF templates obtained
from the QSO, keeping in mind the already mentioned caveat. 
In particular, we used both the background subtracted QSO cropped image and the best
fit QSO analytical profile (this last case should be less contaminated from galaxy
light; however, it has the additional difficulty that the QSO fit is
not perfect, with a noticeable residual left).

Even using these PSF templates, the need for two components to account for the galaxy
light distribution still stands. In particular, the large difference in
orientation between the outer and inner ellipsoids is unchanged.

The residual component is also unchanged. The only appreciable difference is a
slight change in the two S{\'e}rsic indexes; the inner one becomes steeper (0.71),
the outer one shallower (0.57). 
In order to verify if the residual image could have an explanation in
  terms of inefficient removal of bad pixels or cosmic rays, we carefully examined each of the frames that
  compose the summed NACO image. We did not find anything suspicious in none
  of the frames, and therefore we conclude that such 
  interpretations of the residual image are unlikely.

\section{Single image or multiple images}
\label{Single, SIS}

 The zero-th order question that Q0045-3337 and its foreground galaxy pose to us is whether the lensing
 effect is strong, i.e. whether there are other, still unseen, QSO images, and eventually where they should be looked for. 

In fact, if truly there is no image splitting, only a reduced amount
 of information can be extracted on the galaxy (e.g. Narayan \& Schneider
 \cite{narayanschneider1990}, Le Brun et al. \cite{lebrun2000}), though 
 it can be still useful, in our case, to probe the behaviour of the strong lensing
 cross section for a spiral galaxy. 

The simplest galaxy lens model with some physical meaning is the singular
isothermal sphere (SIS). Within the SIS framework, all lens and source quantities are tied by simple analytical relations. In particular, necessary and sufficient condition for multiple images
 formation is that the Einstein radius $\theta_{E}$ has to be greater than half of $\theta_{I}$,
 the distance of the brightest image from the lens ($\theta_{E} \geq 1/2
 \theta_{I}$; Narayan \& Schneider \cite{narayanschneider1990}). The Einstein radius is defined as $\theta_{E} =4\pi \times (\sigma_{\rm v}/c)^2
 \times D_{\rm {ds}}/D_{\rm{s}}$, where $\sigma_{\rm v}$ is the velocity dispersion  of the SIS,
  $D_{\rm{ds}}$ is the angular diameter distances between the source and the
 lens and $D_{\rm{s}}$ is the angular diameter distance between the source and the observer.
The 0.57$''$ discriminating threshold corresponds to a SIS velocity dispersion
of 167 km s $^{-1}$ for a lens galaxy redshift of 0.4, and of 228 km s $^{-1}$
for a redshift of 1. The mass enclosed in $\theta_{I}$ is respectively 7.8
$\times 10^{10}$ M$_{\odot}$ inside 6.1 kpc and 2.2 $\times 10^{11}$ M$_{\odot}$ inside 9.1 kpc.
Since the companion galaxy of Q0045-3337 is a spiral, 
the Tully-Fisher relationship (Pierini \& Tuffs \cite{pierini1999}) can be used to infer a circular velocity range of 218-351 km s $^{-1}$ from the (guessed)
 absolute magnitude.  Under the assumption that the
 $\sigma_{\rm{v}}$ parameter of the SIS mass distribution is 1$ / \sqrt2$ of the  maximum circular
 velocity, the obtained $\sigma_{\rm{v}}$ are 154 km s $^{-1}$ and 248 km s $^{-1}$, and
the inferred Einstein radius ranges from $0.49 ''$ to $0.68 ''$ (Figure
 ~\ref{q0045second}). Such values do not allow us to
 confirm or reject image splitting, keeping in mind the intrinsic dispersion in the Tully-Fisher relationship, the possibility of evolutionary effects 
and in particular the extreme simplification used in assuming a SIS model.

If the lensing is weak, its effects are magnification of point sources and
tangential stretching of extended sources. We will discuss the stretching
effect later.
 In the SIS framework the maximum magnification without image splitting is
 2. Much higher values can be reached, at least in principle, with different
 mass models (Keeton, Kuhlen \& Haiman \cite{keetonetc2005}).

  \section{Multiple images}
\label{multiple}
In this section we investigate the hypothesis that additional, yet unseen (or
not recognized) images are produced. 
We slightly complicate the galaxy model abandoning the spherical
symmetry. Singular and non-singular Isothermal Ellipsoidal Mass Distribution
(SIEMD, PIEMD) and ellipsoidal potential have been studied in some details in
the last $\sim$ 15 years (e.g. Kassiola \& Kovner \cite{kaskov1993},
Schramm \cite{schramm1990}, Kormann, Schneider \& Bartelmann
\cite{kormann1994}). The structure of the
caustic curves of non-spherical models is changed qualitatively, with the
appearance of the astroid-shaped tangential caustic. The odd image theorem
needs no more infinite demagnification of the central maximum image, and the presence
of two caustics allows the production of up to five images. 
 These models should be more realistic than the SIS oversimplification; however, they still lack ``on the field'' tests for spiral lenses, due to the already mentioned paucity of known cases.

 Our goal in this section is not to make a detailed model of the galaxy mass
 distribution, but only to check if strong, image splitting, lensing
 configurations can be compatible with the constraint that no other image is
 seen in the NAOS-Conica data, and if that is the case to lead the way for
 future observations. To reproduce the qualitative behaviour of different
 cases, we extensively used the Java applet SimpLens (Saha \& Williams \cite{sahawill2003}%;\{http://ankh-morpork.maths.qmw.ac.uk/~saha/astron/lens/}%
). Subsequently, we ran some simple simulations with the Gravlens software
 (Keeton \cite{keeton2001}) to verify if the proposed configurations require a
 non-realistic choice of parameters.
 Table~\ref{table1} shows the models used later in the text. 

%Model I and Model II
% are two SIEMD with different values of the Einstein radius parameter, model III
% is a SIEMD with shear while
% model IV is an exponential mass distribution, that we selected as
% representative of highly flattened mass distributions. In this last case, the cusp caustic pierces the radial caustic originating
% a naked cusp geometry.}

\begin{table}[htb]
\begin{minipage}[t]{\columnwidth}
\caption{GravLens models.
Parameters are, from top to bottom, SIEMD Einstein radius parameter,
exponential disk central surface density, ellipticity, position angle, shear strength,
shear angle and exponential disk elliptical scale radius.}
\label{table1}
\centering
\renewcommand{\footnoterule}{}  % to avoid a line before footnotes
\begin{tabular}{lcccc}
\hline \hline
Parameter & Model I & Model II &  Model III& Model IV\\
 & & &  \\      
\hline      

Type & SIEMD & SIEMD &SIEMD$_\gamma$& ExpDisk\\      
 & & &  \\      
\hline      
      
$R_{E,gal}$\footnote{semiminor axis} ($''$)&0.86&0.49&0.86&-- \\       
$\kappa_{0}$\footnote{in units of critical density} & --&-- &--&2.45 \\      

$e$\footnote{defined as (1-q)/q, with q axis ratio} &0.34&0.34 &0.30 &0.67 \\
 
$PA$\footnote{North over East} (deg)&-13 &0 &-5 &-5\\

$\gamma$ &-- &--&0.06&-- \\       
$\phi_{\gamma}$$^d$ (deg)&--&-- &-20&-- \\            
$R\rm{_d}$($''$)& --&-- &--&0.5 \\
\hline
\end{tabular}
\end{minipage}
\end{table}

\subsection{Three images}
\label{three}
\begin {figure*}[htb]
\centering
  \includegraphics[width=17 cm]{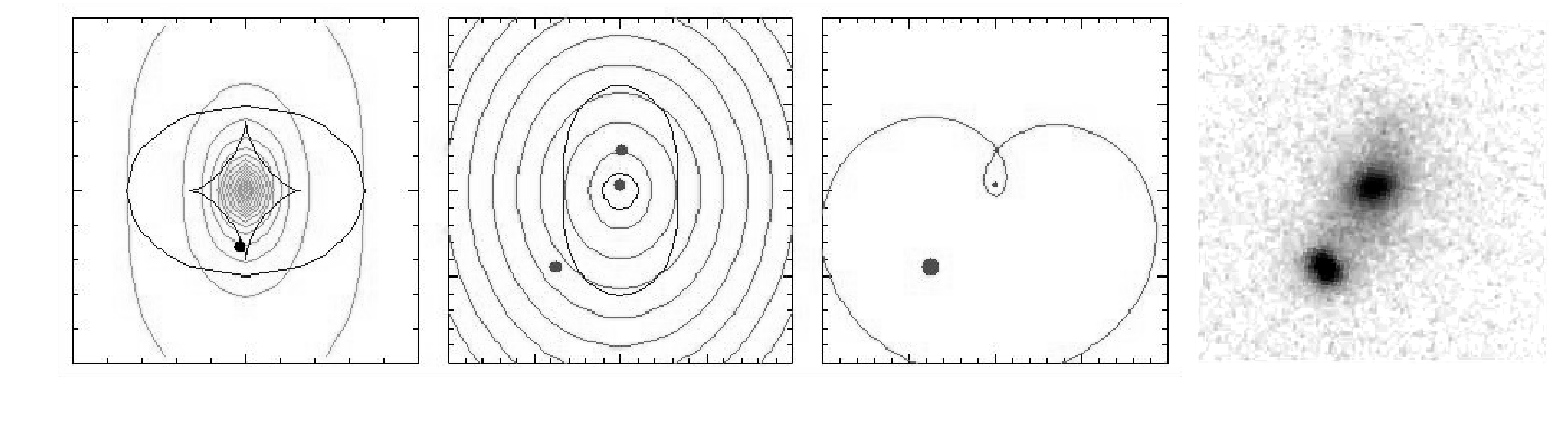}

  \includegraphics[width=17 cm]{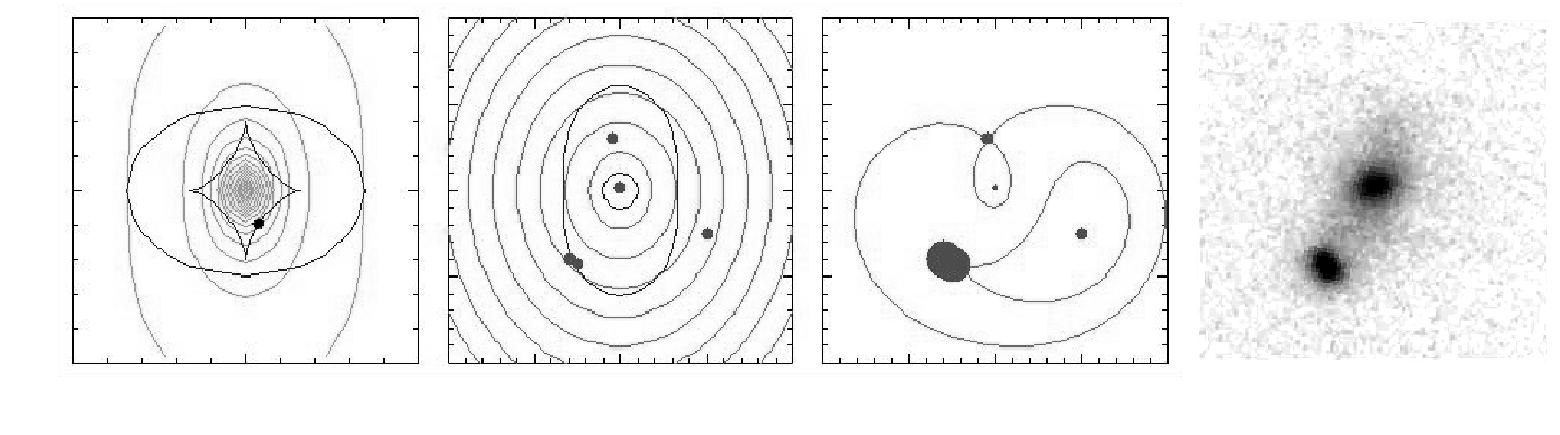}
   
  \caption[Lensing scenarios for Q0045-3337]{\textbf{Lensing scenarios for Q0045-3337.} A qualitative
  representation of lensing scenarios for Q0045-3337. From up to down: three
  images case with the saddle point image hidden; five images case
  with a saddle point image merged with a minimum image, the other
    saddle point and the other minimum images hidden. From left to right: SimpLens source position
  (point), isodensity contours (light curves) and caustics (grey curves);
  SimpLens images positions (points) and critical curves (grey curves);
  SimpLens image positions (points) with area proportional to the
  magnification and virtual light-travel time contours (grey curves);
  Q0045-337 and foreground galaxy, properly rotated and scaled to be directly
  comparable with the SimpLens images. The SimpLens model used is a
  non-singular isothermal ellipsoidal potential, with a core radius of 0.01 and an ellipticity of 0.37.}
  \label{q0045simplens}
\end {figure*}

The easiest, and therefore favoured by the Ockham's razor, image splitting
configuration is a three image configuration (like the one shown in Figure
~\ref{q0045simplens}, top panels). In this case, the source has to be positioned between the external,
radial, caustic and the internal astroid. The maximum, demagnified image is
hidden by the lens galaxy bulge.
Let us suppose that the residual image found in Section ~\ref{surface} is unrelated. 
If this is the case, the absence of other point sources
requires a large magnification ratio between the minimum and the saddle point
images. 
As a working hypothesis we chose a magnification ratio value of 20 as
  threshold of validity for the models. Even if this threshold would be too
  low to be compatible with the absence of any visible trace of the saddle point image, there are other ways to boost the magnification ratio.
In particular, the saddle point image could be hidden by a strong obscuration
 if there is enough dust in the galaxy (even if no evidence of a dust
 lane is found from the surface photometry), or the minimum image flux could be
 enhanced by substructure lensing, microlensing and by the QSO variability itself.

A large magnification ratio can be obtained with a SIEMD in two
  ways (see Keeton, Kuhlen \& Haiman
   \cite{keetonetc2005} for an extensive discussion). First, you can put the source in one
of four drop shaped areas just outside the astroid cusps (see
Figure~\ref{simulationplot}; Keeton, Kuhlen \& Haiman
   \cite{keetonetc2005}, Figure 1). We will refer to this case as to the drops configuration.
If the source is in one of the two, very small, drops along the minor axis, the saddle
point is the brightest image. This do not correspond to our case, and would
require an unphysical mass distribution orthogonal to the light distribution.

If instead the source is in one of the two larger drops along the
  major axis, the
  minimum image is greatly magnified, and it forms very close to the
tangential critical curve; thus the mass scale of the system is roughly fixed
by the distance between the minimum image and the galaxy center.  
Model I of
Table~\ref{table1} is an example of a situation of this kind. The ellipticity
is chosen to not differ too much from the light distribution; then the Einstein
radius parameter (minor semiaxis of the tangential critical curve) is settled by the distance between the QSO and the galaxy
center, and the position angle by the need to stay in the drop area.

The other way to obtain a large magnification ratio is to put the
  source close to the radial caustic, in the external grey area of Figure~\ref{simulationplot}. We refer to this case as to the mouth configuration.
In this case, the saddle
  point image is strongly demagnified. The minimum image forms further away
  from the tangential critical curve. Therefore, for a fixed distance between the minimum image and the lens center, the mass scale
  of the system is reduced. The constrain on the position angle is relaxed
  too. Model II of Table~\ref{table1} is an example of this
  scenario. Note that in this case the Einstein radius parameter is at the
  lower end of the SIS range of Section \ref{Single, SIS}, and the mass
  distribution is not aligned with neither of the light distribution components.

A misalignment between mass and light can be
 mimicked by the presence of shear, or be due to the aftermath of a major
 merger; this last hypothesis could be supported by the isophotal twisting and
 core-disk misalignment of the galaxy. In model III of
   Table~\ref{table1} we added a shear component, that can be either due to
   real shear or just a way to simulate an intrinsic complex angular structure
   of the galaxy mass distribution. With this model the
   production of large magnification ratios is easier, so at our 20 threshold the two areas of
   Figure~\ref{simulationplot} are actually merged (e.g. Keeton, Kuhlen \& Haiman
   \cite{keetonetc2005}, Figure 1).

Let us consider instead the hypothesis that the residual image is the saddle point image. 

The elongated shape of the residual seems to contradict this speculation; however, since we are
dealing with an object near to the plate limit, obtained after subtraction of
an extended component and possibly absorbed, and after all we have no firmly
established PSF shape, the morphological evidence can not be
decisive. 
In the same way the position and the faintness of the residual are less than conclusive clues against the
saddle point image interpretation. The face value magnification
ratio between the QSO and the residual is $\sim$ 80; allowing an underestimate of the residual
luminosity of $\sim$ 0.5 magnitudes, as common for faint sources, and another $\sim$ 0.5 magnitudes of
absorption, we obtain $\sim$ 40.  If indeed the residual feature marks the 
saddle point image, it is not possible to reproduce the positions of
  images and the required magnification ratio with a SIEMD (though this is not
  necessarily true for more
  complex models). Note that in this case absorption is probably less effective in boosting
the magnification ratio, since the saddle point image is not so close to the galaxy
core.

We envisaged also the possibility that the object C is an additional
image of the quasar. It
seems that somehow this
interpretation creates more problems than it solves. In fact the system size of
$\sim4.7 ''$ is excessive for a single galaxy lens. Furthermore, fitting
object C in a
configuration like that of Figure ~\ref{q0045simplens}, top panels, would require a
reversed magnification ratio (A would be the most luminous image),
very difficult to produce unless the mass is orthogonal to light.
Finally, it is not
clear if object C is truly point-like, since it seems to have a slightly
smeared light profile, and the uncertainty on the PSF prevents us from a definite
conclusion in this sense.  

In the three image configuration, the elongation of the quasar has to be of instrumental origin, as is explained in the following subsection.

\subsection {Quasar ellipticity}
\label{ellipticity}
 The first, most obvious, explanation for the ellipticity of Q0045-3337 in
 the NACO image is instrumental, i.e. due to adaptive optics PSF deformation (see
~\ref{surface}).
 As noted by Falomo et
 al. (\cite{fal2005}), Q0045-3337 is elongated in a direction almost aligned
 to the vector radius of the guide star, but not equal (with a difference of
 $\sim$ 10-15 degrees). The galaxy itself, and its brightest core in
 particular, do not appear to suffer similar deformation, but since their
 intrinsic brightness distribution is not radially symmetric, no definite
 conclusion can be drawn. If the ellipticity is real and not an instrumental
 artifact, a gravitational lensing effect can be invoked. In principle,
 image deformation can be produced by lensing of resolved sources, since in
 that case lensing conserves surface brightness, stretching the sources in the
 tangential directions and originating the well known arclets. However, it
 seems unlikely that this can be the case for Q0045-3337. In fact, while we
 stress one last time that the PSF in an adaptive optical image can be
 intrinsically complex, a magnified point source superimposed to a
 tangentially stretched extended source (like a QSO with its host galaxy)
 should result in a sharp bright core with a faint extended wing, i.e. the
 image should appear more elliptical at low brightness, and less at the
 brightness peak, where it should be dominated by the point source. This is
 exactly the opposite of what we see for Q0045-3337 (see the contour map of Figure
~\ref{q0045second}). The only way in which gravitational lensing can produce
such a behaviour is by means of merging two or more images, unresolved by the
NACO observation (note that the merging of two equal circular PSFs can produce an elliptical PSF with stronger ellipticity at the peak and weaker at the periphery). In the following subsection we will explore in some more detail this possibility.

\subsection {Five images}
If we require a lensing explanation to the quasar ellipticity, we need to produce five images. This can be obtained
 in any model with two at least partially nested caustic curves. For our
 purpose, we want the merging of two (or even three) images near to the
 tangential critical curve, in such a way that the NACO observation is not
 sufficient to resolve the contribution of the different images (as in Figure
~\ref{q0045simplens}, bottom panels). 
Images so close to a critical curve are strongly magnified. This can provide a
 explanation to the absence of the other images in the NACO
 observation based solely on strong lensing.

 In particular, it is crucial to explain the non detection of
 a minimum image, that is foreseen quite far away from the galaxy and therefore can not
 be obscured. A conservative view assumes that the missing minimum
   image is fully undisturbed and therefore the full limit of the plate can be reached. We
   estimated a plate limit of 22.8 using the barely
 visible 22.6 residual object. In that case the obtained magnification ratio between the merged
 images and the missing minimum image is $\sim 100$. A less dramatic
 magnification ratio is allowed taking into account absorption and
 the effect of residual galaxy luminosity, and again an underestimate of object D
 magnitude. In particular we educatedly guessed a $\sim 40$ value. 
Note that the centroids of the merged images
 have to be very close each other; to give an order of magnitude example, the plate scale
 of the NACO image is only $0.054 ''$, and the distance between the two components used to fit the QSO
 with GALFIT in Section ~\ref{surface} is $0.052 ''$. 

We verified the viability of the five images scenario numerically with
 Gravlens simulations.

It is possible to obtain such a configuration with a SIEMD at two
  conditions: first, the tangential critical line must cross the position of
  the merged images (thus setting the galaxy mass scale), second, the source
  needs to be very close to the astroid caustic fold (see Figure~\ref{simulationplot}).
Assuming fulfilled the first condition, like in Model I, the source plane area
allowed by the second condition is very small. 
 This result can be improved using different models of mass distribution.
In particular, we employed an highly flattened distribution (so to obtain a
naked cusp geometry), like the exponential disk mass distribution of Model IV.

Although such a strong flattening can barely represent the observational
parameters that we
 have, the simulation proves that in principle this model can produce
 more easily five images with the required  magnification ratio.

\subsection {Five images versus three images}
\begin {figure*}[htb]
\centering
   \includegraphics[height=7.62 cm]{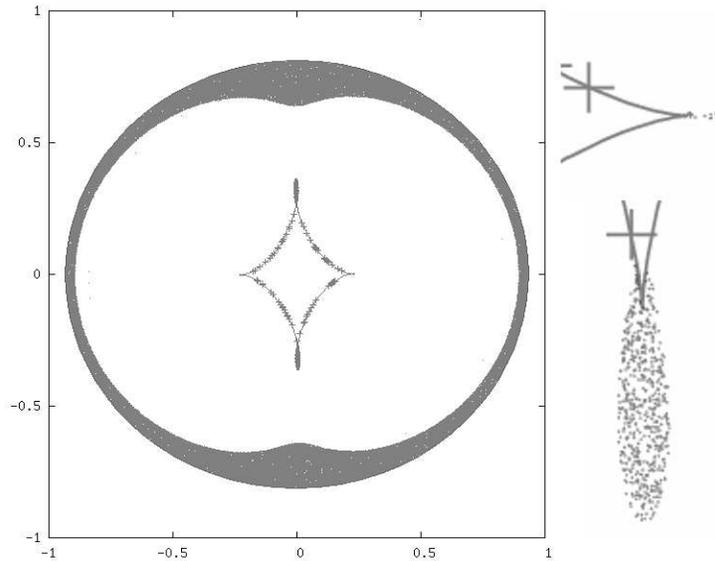}
  
  \caption[simulationplot]{\textbf{Simulated sources.} Source plane of a SIEMD
    model (see Table~\ref{table1} for parameters) with simulated sources. The
    grey area is the locus of sources producing three images with a
    magnification ratio $>$20, crosses
    are sources producing five images with a magnification $>$40. On the
    right, close-ups of two drop areas. More
    details are reported in the text.}
  \label{simulationplot}
\end {figure*}

We evaluated the relative probability of the five images scenario in
  respect to the three images scenario in a statistical way. We chose a galaxy mass model able to reproduce either the three or the five
  images changing the source position. Therefore, the relative probability of the two
  scenarios can be evaluated as a ratio of weighted cross-sections, where the
  weight is needed to keep into account properly the magnification bias. 
We simulated one million source positions for model I with the mock1 GravLens
command (see Figure~\ref{simulationplot}). We then picked up only the realizations able to exceed a fixed magnification
ratio threshold, and we integrated the area spanned, weighting each source realization with its own
magnification bias (e.g. in the integrand). For the three images scenario we
used a 20 threshold of magnification ratio, defined as ratio between the fluxes of the two
brightest images. For the five images scenario, we defined the magnification
ratio as the sum of the fluxes of the two brightest images over the flux of the
third brightest image, and we chose the (optimistic) value of 40 for the threshold. As quasar number counts function we used
a power law with two different values of the index, 1.85 and 2.23 (taken from
the literature, e.g. Richards et al. \cite{richards2005}). The obtained relative probability is respectively 2.44$\%$ and 13.2$\%$.
As expected (it appears as an exponent) the impact of the index of the
quasar number counts function is very strong. Such probability values are not sufficiently low 
to firmly exclude the five images scenario on the basis of
this calculation alone. In fact
while sources producing the five images configuration are rare (look
again at Figure~\ref{simulationplot}), their
overall magnification is large, so the inclusion of the magnification bias
strongly enhances their weight.

The relative probability of the three and five image cases
does not keep into account the distinction between drops and
mouth configurations; in fact to evaluate such a probability as a ratio between
weighted cross-sections, we must make the hypothesis that the galaxy models is
fixed; otherwise, we should evaluate the relative probability of the different
galaxy models, something that goes beyond the scope of the paper. However, it
is important to make clear that while the five images case requires
strictly such a (rather large) value of the Einstein radius parameter, since the merged
images are on the two sides of the tangential critical curve, with the three
images it is allowed a wider range
of values. In particular, in the mouth configuration, it is possible to reproduce the
observational constraints with a much lower value of the Einstein radius
parameter. Such a smaller Einstein radius
would be favoured both by the SIS/Tully-Fisher evaluation of Section \ref{Single, SIS}, and by
the different normalizations proposed for application of the SIEMD to disk
galaxies (e.g. Keeton \& Kochanek \cite{keetonkochanek1998}). Furthermore, the mouth configuration leaves much more room for
explaining the absence of the saddle point image, since it puts it much closer
to the galaxy core.

\section{Discussion}
\label{discussion}
We performed two dimensional photometry on the galaxy near to Q0045-3337. We
confirmed its spiral nature and found evidence for two components with
different orientations. We found a residual image after model subtraction,
that is most likely due to a spiral arm, even if more interesting
possibilities can not be excluded.
We then verified that no unusual parameters for the galaxy are required to
produce, or not produce, image splitting. We also found some very simple strong lensing
configuration capable of not contradicting existing data, either
assuming a) that
no other image is seen and Q0045-3337 observed elongation is instrumental, b)
that the residual image is truly another lensed image of Q0045-3337, and still
the elongation is instrumental, c) that Q0045-3337 elongation is due to the
merging of two or more images, and no other image is seen. We
  evaluated the likelihood of the different cases, finding that 
case a) is the most probable, even if cases b) and c) can not be rejected.

All these speculations can be easily verified observationally.  In particular,
another optically adaptive image with more information on the PSF behaviour
would confirm or rule out the ellipticity of Q0045-3337 and 
 en passant it would also verify if object C is point-like. A measurement of the galaxy
 redshift, while probably not conclusive, would help to constrain how likely
 is the production of multiple images. Deeper pointings could reveal
 additional images or put tighter constraints on required magnification
 ratios. 
If case c) turns out to be the most probable, an HST image could resolve the merged images.
At present, strong lensing effects on Q0045-3337 can only be labelled as
possible. 
Spiral strong lenses are rare objects. The current view assumes that the
cross-section for image splitting is not generally increased  by the
presence of a thin massive disk, at least once corrected for the inclination and
once properly taken in account the current observational capabilities;
furthermore, absorption biases strongly against the detection (Bartelmann \&
Loeb \cite{bartelmannloeb1998}; Keeton \& Kochanek \cite{keetonkochanek1998}; Perna, Loeb \& Bartelmann \cite{perna1997}).  
In such a context, the galaxy close to Q0045-3337 can offer valuable help to
our knowledge of spiral galaxy mass distributions, and to lensing statistics
too. Even in the case without image splitting, as it is likely both for the
sample of Narayan \& Schneider
 (\cite{narayanschneider1990}) and for the Lyman absorbers of Le Brun et
 al. (\cite{lebrun2000}), interesting upper limits can be placed on the galaxy
 mass inside the Einstein radius (once the redshift is known). Furthermore,
 the lensing effect of stretching of resolved sources could make Q0045-337
 again a valuable target for deeper host galaxy studies (as it was originally
 selected by Falomo et al.(\cite{fal2005}).  
As a final comment, it is interesting to note that this case can be a perfect
example of advantages and drawbacks of adaptive optics applied to strong lensing.

 In the last
years, HST has been the principal instrument in strong lensing studies, and it is
interesting to evaluate if it can be -at least partially- surrogated by adaptive optics 
earth telescopes, especially in the case of a gap between the end of HST
operations and
the start of the James Webb Space Telescope operations.
As Q0045-3337 and its neighbour galaxy tell us, a necessary condition to be
fulfilled in this case is a correct PSF evaluation observational procedure
(as was at first proposed
for  Q0045-3337); otherwise, the unresolved uncertainty between real and
instrumental deformations hampers the
exploitation of the optimal resolution for the purpose of lens studies.

\begin{acknowledgements}
  We thank Aldo Treves and Renato Falomo for pointing out to us the existence of
  Q0045-3337, for useful discussions on lensing or instrumental origin of the
  observed deformation, and for kindly sharing with us the NAOS-CONICA ESO VLT
  observational data.   
We thank the referee, Neal Jackson, for his substantial contribution to the
improvement of the paper.  
Marco Miranda is supported by the Swiss National Science Foundation.
\end{acknowledgements}


\begin{thebibliography}{}

\bibitem[2006]{adel2006}
Adelman-McCarthy, J., Agueros, M.~A., Allam, S.~S., et al. 2006, ApJS, 162, 38

\bibitem[1998]{bartelmannloeb1998}
Bartelmann, M., \& Loeb, A. 1998, ApJ, 503 48

\bibitem[2003]{browne2003}
Browne, I.~W.~A., Wilkinson, P.~N., Jackson, N.J.F., et al. 2003, MNRAS, 341, 13

%\bibitem[2006]{castander06}
%Castander, F.~J., Treister, E., Maza, J. \& Gawiser, E. 2006, ApJ accepted
%and in press (scheduled Dec 1 2006, 652, 955), astro-ph 061130

\bibitem[2006]{clenet2006}
 Cl{\'e}net, Y., et al. 2006, in Proc. SPIE Vol. 6272, Advances in
 Adaptive Optics II, ed.\ Brent L. Ellerbroek, \& Domenico Bonaccini Calia, 62723T

\bibitem[2005]{fal2005}
Falomo, R., Kotilainen, J.~K., Scarpa, R. \& Treves, A. 2005, A\&A, 434, 469

\bibitem[1985]{huchra1985}
Huchra, J., Gorenstein M., Kent, S., et al. 1985, AJ, 90, 691

\bibitem[1996]{iovino1996}
Iovino, A., Clowes, R. \& Shaver, P. 1996, A\&AS, 119, 265

\bibitem[1995]{jackson1995}
Jackson, N., de Bruyn, A.~G., Myers, S., et al. 1995, MNRAS, 274, L25

%\bibitem[1998]{jackson1998}
%Jackson, N., Nair, S., Browne I.~W.~A., et al. 1998, MNRAS, 296, 483

\bibitem[1993]{kaskov1993}
Kassiola, A., \& Kovner, I. 1993, ApJ, 417, 450

\bibitem[2001]{keeton2001}
Keeton, C.~R. 2001, ApJ submitted, astro-ph/0102340 

\bibitem[1998]{keetonkochanek1998}
Keeton, C.~R., \& Kochanek, C.~S. 1998, ApJ, 495, 157

\bibitem[2005]{keetonetc2005}
Keeton, C.~R., Kuhlen, M. \& Haiman, Z. 2005, ApJ, 621, 559

\bibitem[2004]{diablerets}
Kochanek, C.~S., Schneider, P. \& Wambsganss, J. 2004, Part 2 of Gravitational
Lensing: Strong, Weak \& Micro, Proceedings of the 33rd Saas-Fee Advanced Course, ed.\ G. Meylan, P. Jetzer \& P. North (Springer-Verlag: Berlin)

\bibitem[1994]{kormann1994}
Kormann, R., Schneider, P. \& Bartelmann, M. 1994, A\&A, 284, 285

\bibitem[2000]{lebrun2000}
Le Brun, V., Smette, A., Surdej, J. \& Claeskens, J.-F. 2000, A\&A, 363, L837

%\bibitem[2002]{lewis2002}
%Lewis, G.~F., Ibata, R.~A., Ellison S.~L., et al. 2002, MNRAS, 334, L7

\bibitem[2003]{myers2003}
Myers, S.~T., Jackson, N.~J., Browne, I.~W.~A., et al. 2003, MNRAS, 341, 1

\bibitem[1990]{narayanschneider1990}
Narayan, R., \& Schneider, P. 1990, MNRAS, 243, 192

\bibitem[1993]{patnaik1993}
Patnaik, A.~R., Browne, I.~W.~A., King, L.~J., et al. 1993, MNRAS, 261, 435

\bibitem[2002]{peng2002}
 Peng, C.~Y., Ho, L.~C., Impey, C.~D. \& Rix, H.-W. 2002, AJ, 124, 266

\bibitem[1997]{perna1997}
Perna, R., Loeb, A., \& Bartelmann, M. 1997, ApJ, 488, 550

\bibitem[1999]{pierini1999}
Pierini, D., \& Tuffs, R.~J. 1999, A\&A, 343, 751

\bibitem[1988]{prs1988}
Pramesh Rao, A., \& Subrahmanyan, R. 1988, MNRAS, 231, 229

\bibitem[2005]{richards2005}
Richards, G.~T., Croom, S.~M., Anderson, S.~F., et al. 2005, MNRAS, 360, 839

\bibitem[2003]{sahawill2003}
Saha, P., \&  Williams, L.~L.~R. 2003, AJ, 125, 2769

\bibitem[1990]{schramm1990}
Schramm, T. 1990, A\&A, 231, 19

\bibitem[2001]{veronveron2001}
V\'{e}ron-Cetty, M.~P., \& V\'{e}ron, P. 2001, A\&A, 374, 92 

\bibitem[2003]{winn2003}
Winn, J.~N., Hall, P.~B., \& Schechter, P.~L. 2003, ApJ, 597, 672

\bibitem[2000]{white1997}
White, R.~L., Becker, R.~H., Helfand, D.~J., et al. 1997, ApJ, 475, 479

\bibitem[2000]{york2000}
York, D.~G., Adelman, J., Anderson, J.~E., et al. 2000, AJ, 120, 1579

\end{thebibliography}
\end{document}